\documentclass[letterpaper,twocolumn,10pt]{article}
\pdfoutput=1
\usepackage{usenix,epsfig,endnotes}
\usepackage{graphicx,subfigure,array}
\usepackage{times,amsmath,amssymb,verbatim}
\usepackage[usenames,dvipsnames]{color}
\usepackage{cite,footnote,verbatim,threeparttable}
\usepackage[font=small,labelfont=bf]{caption}
\usepackage{multirow,algorithm,algorithmic}
\usepackage[hyphens]{url}
\usepackage{breakurl}
\newcommand{\paragraphb}[1]{\vspace{0.05in}\noindent{\bf #1} }
\DeclareMathOperator*{\argmin}{arg\,min}

\ifpdf   \else \fi

\begin{document}

\date{}

\title{\Large \bf Fingerprinting Smart Devices Through Embedded Acoustic Components}

\author{
{\rm Anupam Das, Nikita Borisov} and {\rm Matthew Caesar}\\
University of Illinois at Urbana-Champaign
} 

\maketitle

\thispagestyle{empty}

\subsection*{Abstract}

The widespread use of smart devices gives rise to both security and privacy concerns. Fingerprinting smart devices can assist in
authenticating physical devices, but it can also jeopardize privacy by allowing remote identification without user awareness. We propose a
novel fingerprinting approach that uses the microphones and speakers of smart phones to uniquely identify an individual device. During fabrication, subtle imperfections arise in
device microphones and speakers which induce anomalies in produced and received sounds. We exploit this observation to fingerprint smart devices
through playback and recording of audio samples. We use audio-metric tools to analyze and explore different acoustic features and analyze their
ability to successfully fingerprint smart devices. Our experiments show that it is even possible to fingerprint devices that have the same vendor and model; we were able to accurately distinguish over 93\% of all recorded audio clips from 15 different units of the same model.
Our study identifies the prominent acoustic features capable of fingerprinting devices with high
success rate and examines the effect of background noise and other variables on fingerprinting accuracy.

\section{Introduction}

Mobile devices, including smartphones, PDAs, and tablets, are quickly becoming widespread in modern society. In 2012 a total of 1.94 billion mobile
devices were shipped, of which 75\% were smart and highly-featured phones~\cite{canalys,mobithink,gartner}.
Canalys predicted that the mobile device market will reach 2.6 billion units by 2016, with smartphones and tablets continuing to
dominate shipments~\cite{canalys} .
The rapid uptake of intelligent mobile devices is not surprising, due to the numerous advantages they provide consumers,
from entertainment and social applications to business and advanced computing capabilities. However, smartphones, with all their interactive,
location-centric, and connectivity-based features impose threatening concerns on user privacy and information security. There has been a large body
of research work highlighting and proposing solutions for privacy and security issues of
smartphones~\cite{privacybreach,appattack,egele11:pios,Gibler:2012,Enck:2010,Shabtai:2010,zhou:2011}.
All these works center around securing the software, including the operating system and network stack, of mobile devices, for example by instilling
fine-grained access control policies, or restricting dataflow, containing private data, to a network sink.


In this paper we propose a novel technique for fingerprinting the {\em hardware} of smartphones.
The observation is that even if the software on mobile devices is strengthened, hardware-level idiosyncrasies in microphones and speaker can be used to fingerprint physical devices.
During manufacturing, imperfections are introduced in the analog circuitry of these components, and as such, two microphones and speakers are never alike.
Through an observational study, we find that these imperfections are substantial enough, and prevalent enough, that we can reliably distinguish
between devices by passively observing audio, and conducting a simple spectral analysis on the recorded audio.
Our approach can substantially simplify the ability for an adversary to track and identify people in public locations, identify callers, and produce other
threats to the security and privacy of mobile device users. Our approach works well even with few samples --- for example, we show that with our
techniques, an adversary could even use the short ringtones produces by mobile device speakers to reliably track users in public environments.

Our approach centers around development of a novel fingerprinting mechanism, which aims to ``pull out'' imperfections in device circuitry. Our
mechanism has two parts: a method to extract auditory fingerprints and a method to efficiently search for matching fingerprints from a database. To
generate fingerprints of speakers we record audio clips played from smartphones on an external device (i.e., laptop/PC) and vice versa for generating
fingerprints of microphones. We use two different classifiers to evaluate our fingerprinting approach. Moreover, we test our fingerprinting approach
for different genre of audio clips at various frequencies. We also elaborately study various audio features that can be used to accurately
fingerprint smartphones. Our study reveals that mel-frequency cepstral coefficient (MFCC) is the dominant feature for fingerprinting smartphones. We
also analyze the sensitivity of our fingerprinting approach against different factors like sampling frequency, distance between speaker and recorder,
training set size and ambient background noise.

\paragraphb{Contributions.} We offer the following contributions:\nolinebreak
\begin{itemize}
\setlength{\itemsep}{1pt}
\setlength{\parskip}{0pt}
\setlength{\parsep}{0pt}

\item [$\bullet$] We propose a novel approach to fingerprinting smart devices. Our approach leverages the manufacturing idiosyncrasies of microphones
and speakers embedded in smart devices.

\item [$\bullet$] We study feasibility of a spectrum of existing audio features that can be used to accurately fingerprint smartphones. We find that
the mel-frequency cepstral coefficient (MFCC) performs particularly well for fingerprinting smartphones.

\item [$\bullet$] We investigate two different classifiers to evaluate our fingerprinting approach. We conclude that Gaussian Mixture Models
(GMM) are more effective in classifying our recorded audio fingerprints.

\item [$\bullet$] We perform experiments across several different genres of audio excerpts. We also analyze how different factors like sampling frequency,
distance between speaker and recorder, training set size and ambient background noise impact the accuracy of our fingerprinting.

\item [$\bullet$] Finally, we discuss how our fingerprinting approach can be used as an additional factor for authentication.
\end{itemize}

\paragraphb{Roadmap.} The remainder of this paper is organized as follows. Section {\ref{overview}} gives an overview of our fingerprinting
approach. We discuss why microphones and speakers integrated in smartphones can be used to generate unique fingerprints in Section
{\ref{microphone-speaker}}. Section {\ref{feature_algo}} describes the different audio features considered in our experiments, along with the
classification algorithms used in our evaluation. Section {\ref{evaluation}} elaborately presents our experimental results. We discuss two diametric
applications of our device fingerprinting in Section {\ref{applications}}. We describe some related works in Section {\ref{related-work}}. Section
{\ref{limitations}} discusses some limitations of our approach. Finally we conclude in Section {\ref{conclusion}}.

\section{Overview}{\label{overview}}


In this section we give an overview of our approach, and identify the key challenges that we address in this paper.

The key insight behind our work is that imperfections in smart device hardware induce unique signatures on received/transmitted audio, and these
unique signatures, if identified, can be used to fingerprint the device. Our approach consists of three key components. The first challenge we
encounter is acquiring a set of audio samples for analysis in the first place. To do this, we have a {\em listener} module, responsible for receiving
and recording device audio. The listener module could be deployed as an application on the smart device (many mobile OSes allow direct access to
microphone inputs), or as a stand alone (e.g., the adversary has a microphone in a public setting to pick up device ringtones). The next challenge is
to effectively identify device signatures from the received audio stream. To do this, we have an {\em analyzer} module, which leverages signal
processing techniques to localize spectral anomalies, and construct a `fingerprint' of the auditory characteristics of the device.

\begin{figure}[!htp]
\centering
\includegraphics[width=1.0\columnwidth]{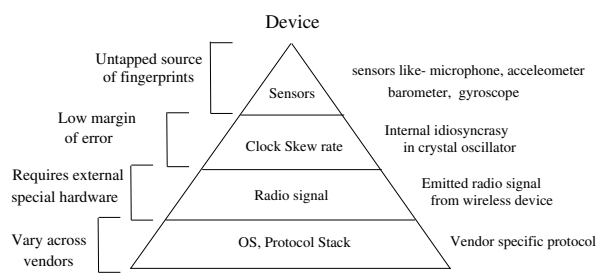}
\caption{Device specific features that can be exploited to uniquely distinguish devices.} \label{device-biometrics}
\end{figure}

A key question that remains, which forms a major focus of this paper, is in construction of an effective fingerprinting
scheme. Our goal is to determine a scheme that maximizes the ability to distinguish different devices.
To do this, it helps to have some understanding of how devices differ at a physical level.
Devices can vary at different layers of the manufacturing process. The most obvious way to distinguish devices manufactured by different vendors is
to analyze the protocol stack installed in the devices. Usually different vendors have their own distinct features integrated inside the protocol
stack. A close analysis of the protocol stack can help in distinguishing devices from different vendors. However, this approach is not helpful in
distinguishing devices produced by the same vendor. To distinguish devices produced by the same vendor we need to look more deeply into the devices
themselves because at the hardware level no two device are same. Hardware imperfections are likely to arise during the manufacturing process of
sensors, radio transmitters and crystal oscillators suggesting the existence of unique fingerprints. This idiosyncrasies can be exploited to
distinguish devices. Figure \ref{device-biometrics} illustrates the different device specific features that could be utilized to identify devices
uniquely.  We investigate properties of device hardware in more detail in Section~\ref{microphone-speaker}.

A second aspect to this question is what sort of audio analysis techniques are most effective in identifying unique signatures of device hardware.
There are a large number of audio properties which could be used (spectral entropy, zero crossings, pitch, etc.) as well as a broad spectrum of
analysis algorithms that can be used to summarize these properties (principle component analysis, linear discriminant analysis, feature selection,
etc.).  We will study alternative properties to characterize hardware-induced auditory anomalies in Section~\ref{audio-feature} as well as algorithms
for effectively clustering them in Section~\ref{classification-algo}.


\section{Source of Fingerprints}{\label{microphone-speaker}}

In this section we will take a closer look at the microphones and speakers embedded on today's smartphones. This will help understand how
microphones and speakers can act as a potential source of unique fingerprints.

\subsection{Closer Look at Microphones}
\begin{figure*}[!htb]
\centering
\begin{tabular}{lll}
\raisebox{+0.5\height}{\includegraphics[width=0.35\columnwidth]{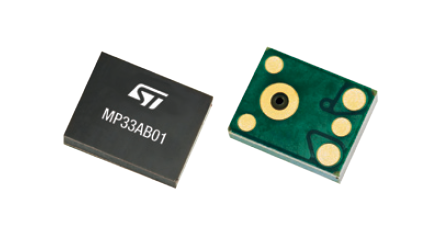}}&\raisebox{1.5cm}{\Large{$\Rightarrow$}}&\includegraphics[
width=1.5\columnwidth]{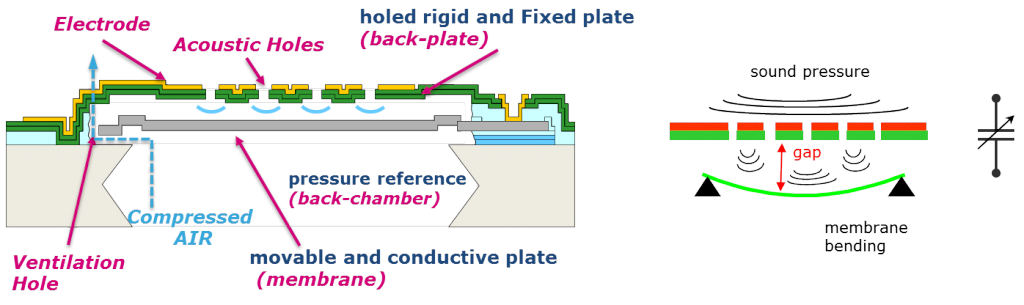}
\end{tabular}
\caption{The internal architecture of MEMS microphone chip used in smartphones.} \label{microphone}
\end{figure*}

Microphones in modern smartphones are based on Micro Electro Mechanical Systems (MEMS)~\cite{microphone1,microphone2,microphone3}. To enhance active
noise and echo canceling capabilities, most smartphones today have more than one MEMS microphone. For example, the iPhone 5 has a total of three
embedded MEMS microphones~\cite{microphone2}. According to the IHS-iSuppli report, Apple and Samsung were the top consumers of MEMS microphones in
2012, accounting for a combined 54\% of all shipped MEMS microphones~\cite{microphone1}.

A MEMS microphone, sometimes called a microphone chip or silicon microphone, consists of a coil-less pressure-sensitive diaphragm directly etched
into a silicon chip. It is comprised of a MEMS die and a complementary metal-oxide-semiconductor (CMOS) die combined in an
acoustic housing~\cite{microphone-model,howMEMSworks}.
The CMOS often includes both a preamplifier as well as an analog-to-digital (AD) converter.
Modern fabrication techniques enable highly compact deigns, making them
well suited for integration in digital mobile devices. The internal architecture of a MEMS
microphone is shown on Figure~\ref{microphone}. From the figure we can see that the MEMS microphone's physical design is based on a variable capacitor consisting
of a highly flexible diaphragm in close proximity to a perforated, rigid back-plate. The perforations permit the air between the diaphragm and
back-plate to escape. When an acoustic signal reaches the diaphragm through the acoustic holes, the diaphragm is set in motion. This mechanical
deformation causes capacitive change which in turn causes voltage change. In this way sound pressure is converted into an electrical signal for further
processing. The back-chamber acts as a acoustic resonator and the ventilation hole allows the air compressed inside the back chamber to flow out,
allowing the diaphragm to move back into its original place.

The sensitivity of the microphone depends on how well the diaphragm deflects to acoustic pressure; it also depends on the gap between
the static back-plate and the flexible diaphragm. Unfortunately, even though the manufacturing process of these microphones has been streamlined, no two chips
roll off the assembly line functioning in exactly the same way\footnote{Imperfections can arise for the following reasons: slight variations in the
chemical composition of components from one batch to the next, wear in the manufacturing machines or changes in temperature and humidity.}. While
subtle imperfections in the microphone chips may go unnoticed by human ears, computationally such discrepancies may be sufficient to discriminate
them, as we later show.

\subsection{Closer Look at Microspeakers}
\begin{figure*}[!htb]
\centering
\begin{tabular}{ccc}
\raisebox{+0cm}{\includegraphics[width=0.6\columnwidth]{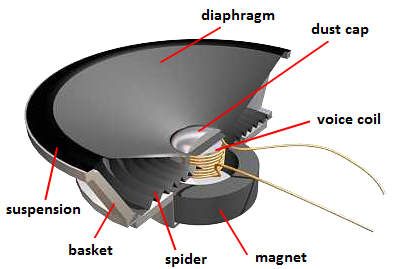}}&\raisebox{+0.4\height}{\includegraphics[width=0.35\columnwidth]{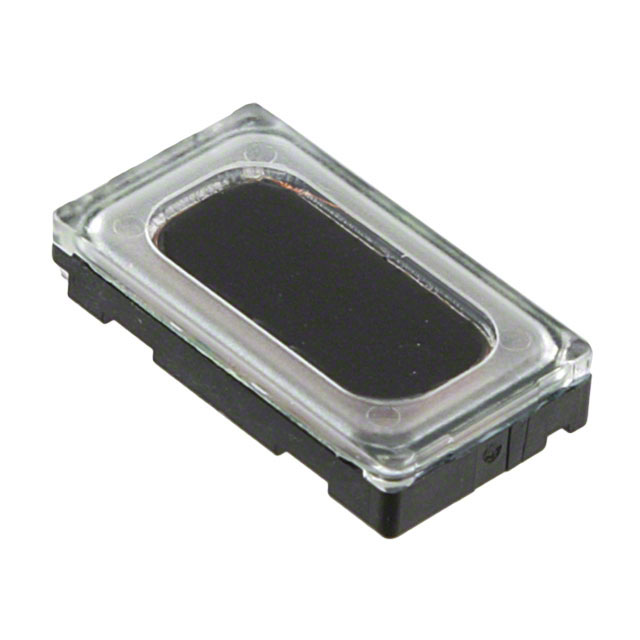}}
&\includegraphics[width=0.6\columnwidth]{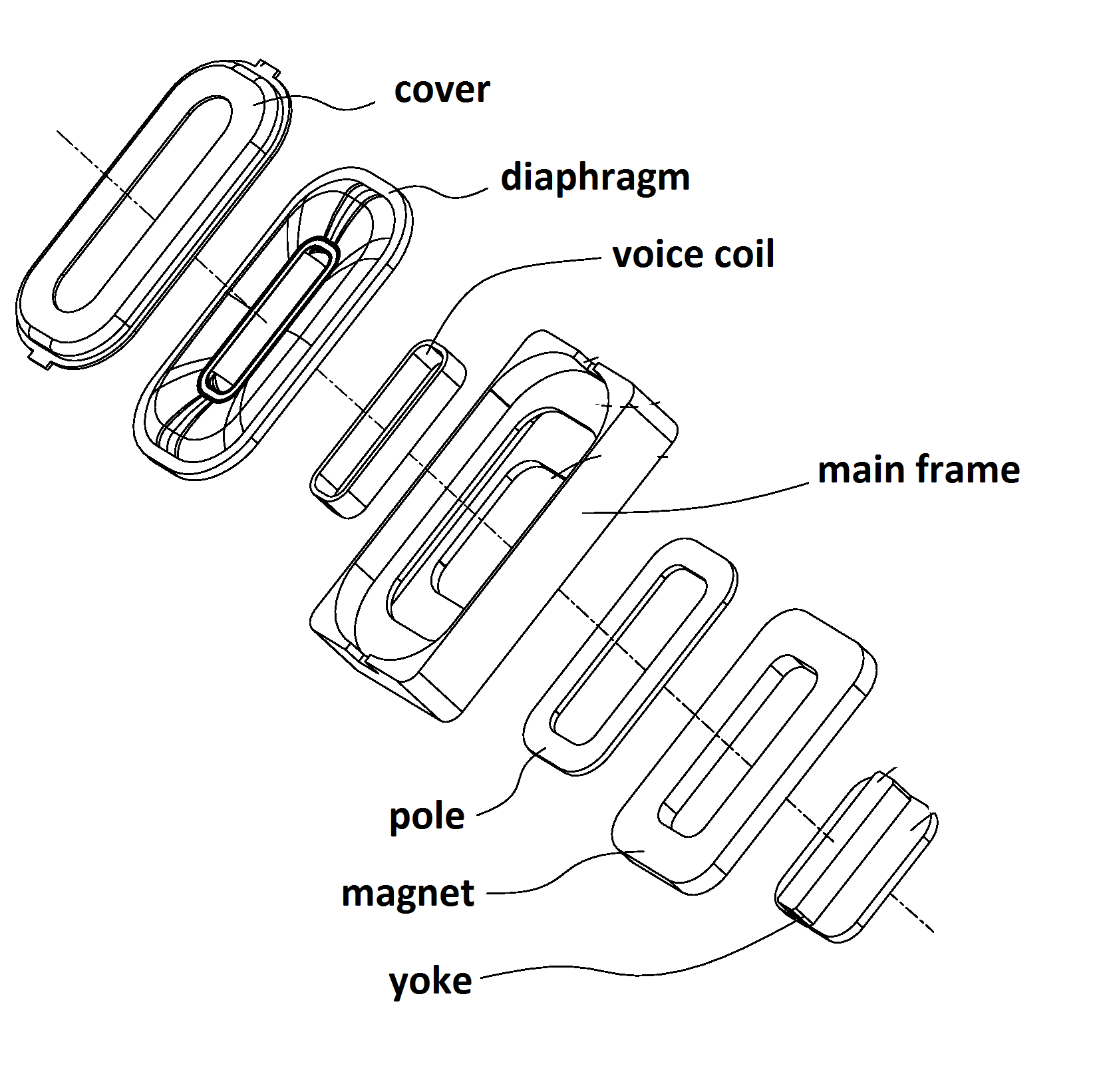}\\
(a)&(b)&(c)
\end{tabular}
\caption{(a) The basic components of a speaker, (b) A typical MEMS microspeaker used in smartphones, (c) The internal architecture of a
microspeaker chip.} \label{microspeaker}
\end{figure*}

Micro-speakers are  a scaled down version of a basic acoustic speaker. So lets first look at how speakers work before we discuss how
microspeakers can be used to generate unique fingerprints. Figure~\ref{microspeaker}(a) shows the basic components of a speaker. The diaphragm is
usually made of paper, plastic or metal and its edges are connected to the suspension. The suspension is a rim of flexible material that allows the
diaphragm to move. The narrow end of the diaphragm's cone is connected to the voice coil. The voice coil is attached to the basket by a spider (damper),
which holds the coil in position, but allows it to move freely back and forth. A permanent magnet is positioned directly below the voice coil.

Sound waves are produced whenever electrical current flows through the voice coil, which acts as an electromagnet. Running varying electrical current
through the voice coil induces a varying magnetic field around the coil, altering the magnetization of the metal it is wrapped around. When the
electromagnet's polar orientation switches, so does the direction of repulsion and attraction. In this way, the magnetic force between the voice coil
and the permanent magnet causes the voice coil to vibrate, which in turn vibrates the speaker diaphragm to generate sound waves.

Figure~\ref{microspeaker}(b) shows a typical MEMS microspeaker chip and Figure~\ref{microspeaker}(c) shows the components inside the
microspeaker~\cite{zhang2011micro,chang2012speaker}. The components are similar to that of a basic speaker; the only difference is the size and
fabrication process~\cite{Shahosseini:2010,Sang-Soo:2009,Cheng:2004}. The amplitude and frequency of the sound wave produced by the speaker's
diaphragm is dictated respectively by the distance and rate at which the voice coil moves. However, due to the inevitable variations and
imperfections of the manufacturing process, no two speaker are going to be
alike. Thus, subtle differences in sound generated by different speakers can arise. In our work, we develop techniques to computationally localize
and evaluate these differences.

\section{Audio Features and Classification Algorithms}{\label{feature_algo}}
In this section we briefly describe the acoustic features that we used in generating fingerprints. We also discuss the classification algorithms used
in identifying the devices from which the fingerprints originated.

\subsection{Audio Features}{\label{audio-feature}}
Given our knowledge that imperfections exist in device audio hardware, we now need some way to detect them. To do this, our approach identifies {\em
acoustic features} from an audio stream, and uses the features to construct a {\em fingerprint} of the device. Computing acoustic features from an
audio stream is a subject of much  research~\cite{Mckinney03,Cano:2005,Bartsch:2005,Tzanetakis2002}. To gain an
understanding of how a broad range of acoustic features are affected by device imperfections we investigate a total of 15 acoustics features (listed
in Table~\ref{allfeatures}), all of which have been well-documented by researchers. A detailed description of each acoustic feature is available in
Appendix~\ref{appendix}.

\begin{table*}[tbp]
\centering \caption{Explored acoustic features} \resizebox{15.5cm}{!}{
\begin{tabular}{|c|c|c|c|}
\hline
\#&Feature&Dimension&Description\\
\hline
1&RMS&1&Square root of the arithmetic mean of the squares of the signal strength at various frequencies\\
\hline
2&ZCR&1&The rate at which the signal changes sign from positive to negative or back\\
\hline
3&Low-Energy-Rate&1&The percentage of frames with RMS power less than the average RMS power for the whole audio signal\\
\hline
4&Spectral Centroid&1&Represents the center of mass of a spectral power distribution\\
\hline
5&Spectral Entropy&1&Captures the peaks of a spectrum and their locations\\
\hline
6&Spectral Irregularity&1&Measures the degree of variation of the successive peaks of a spectrum\\
\hline
7&Spectral Spread&1&Defines the dispersion  of the spectrum around its centroid\\
\hline
8&Spectral Skewness&1&Represents the coefficient of skewness of a spectrum\\
\hline
9&Spectral Kurtosis&1&Measure of the flatness or spikiness of a distribution relative to a normal distribution\\
\hline
10&Spectral Rolloff&1&Defines the frequency below which 85\% of the distribution magnitude is concentrated\\
\hline
11&Spectral Brightness&1&Computes the amount of spectral energy corresponding to frequencies higher than a given cut-off threshold\\
\hline
12&Spectral Flatness&1&Measures how energy is spread across the spectrum\\
\hline
13&MFCCs&13&Compactly represents spectrum amplitudes\\
\hline
14&Chromagram&12&Representation of the distribution of energy along the 12 distinct semitones or pitch classes\\
\hline
15&Tonal Centroid&6&Maps a chromagram onto a six-dimensional Hypertorus structure\\
\hline
\end{tabular}}
\label{allfeatures}
\end{table*}

\subsection{Classification Algorithms}{\label{classification-algo}}

Next, we need some way to leverage the set of features to perform device identification.
To achieve this, we leverage a {\em classification algorithm},
which takes observations (features) from the observed device as input,
and attempts to classify the device into one of several previously-observed sets.

To do this, our approach works as follows. First, we perform a training step, by collecting a number of observations from a set of devices. Each
observation (data point) corresponds to a set of features observed from that device, represented as a tuple with one dimension per feature. As such,
data points can be thought of as existing in a hyper-dimensional space, with each axis corresponding to the observed value of a corresponding
feature. Our approach then applies a classification algorithm to build a representation of these data points, which can later be used to associate
new observations with device types. When a new observation is collected, the classification algorithm returns the most likely device that caused the
observation.

To do this effectively, we need an efficient classification algorithm. In our work, we compare performance of two alternate approaches
described below: \emph{$k$-nearest neighbors} (associates an incoming data point with the device corresponding to the nearest ``learned'' data points),
and \emph{Gaussian mixture models} (computes a probability distribution for each device, and determines the maximally-likely
association).




\paragraphb{$k$-NN:}
The $k$-nearest neighbors algorithm ($k$-NN) is a non-parametric lazy learning algorithm. The term ``non-parametric'' means that the $k$-NN algorithm does not
make any assumptions about the underlying data distribution, which is useful in analyzing real-world data with complex underlying distribution. The term ``lazy learning'' means that the $k$-NN algorithm does not use the training data to make any generalization, rather all the
training data are used in the testing phase making it computationally expensive (however, optimizations are possible). The $k$-NN algorithm works by
first computing the distance from the input data point to all training data points and then classifies the input data point by taking a majority vote
of the $k$ closest training records in the feature space~\cite{duda2001pattern}. The best choice of $k$ depends upon the data; generally, larger
values of k reduce the effect of noise on the classification, but make boundaries between classes less distinct. We will discuss more about the
choice of $k$ in Section~\ref{evaluation}.

\paragraphb{GMM:}
A Gaussian mixture model is a probabilistic model that assumes all the data points are generated from a mixture of a finite number of Gaussian
distributions with unknown parameters. The unknown patterns and mixture weights are estimated from training samples using an \emph{expectation--maximization}
(EM) algorithm~\cite{Dempster77maximumlikelihood}. During the matching phase the fingerprint for an unknown recording is first compared with a
database of pre-computed GMMs and then the class label of the GMM that gives the highest likelihood is returned as the expected class for the unknown
fingerprint. GMMs are often used in biometric systems, most notably in human speaker recognition systems, due to their capability of representing a large
class of sample distributions~\cite{Reynolds200019,Tzanetakis2002}.

\section{Evaluation}{\label{evaluation}}

In this section we perform a series of experiments to evaluate how well we can fingerprint smartphones  by exploiting the manufacturing
idiosyncrasies of microphones and speakers embedded in them. We start by describing how we performed our experiments (Section~\ref{methodology}).
Next, we briefly discuss the setup for fingerprinting devices through speakers and microphones (Section~\ref{testspeaker} and \ref{testmic}). We then
look at fingerprinting devices made by different vendors (Section~\ref{diffvendor}) and later on focus on identifying devices manufactured by the
same vendor (Section~\ref{samevendor}). We also perform an analysis of which features help most when identifying devices from the same vendor, by
determining the {\em dominant} (most-relevant) set of audio features (Section~\ref{SFfeature}). The performance of our approach is affected by
certain aspects of the operating environment, and we study sensitivity to such factors in Section~\ref{sensitivity}.

\subsection{Methodology}{\label{methodology}}

To perform our experiments, we constructed a small testbed environment with real smartphone device hardware. In particular, our default environment
consisted of a 266 square foot (14'x19') office room, with nine-foot dropped ceilings with polystyrene tile, comprising a graduate student office in
a University-owned building (used to house the computer science department). The room was filled with desks and chairs, and opens out on a public
hall with foot traffic. The room also receives a minimal amount of ambient noise from air conditioning, desktop computers, and florescent lighting.
We placed smartphones in various locations in the room. To emulate an attacker, we placed an ACER Aspire 5745 laptop in the room. To investigate
performance with inexpensive hardware, we used the laptop's built-in microphone to collect audio samples (an attacker willing to purchase a
higher-quality microphone may attain better performance). We investigate how varying this setup affects performance of the attack in
Section~\ref{sensitivity}.

\paragraphb{Devices and tools:}
We tested our device fingerprinting on devices from five different manufacturers. Table~\ref{phones} highlights the model and quantities of the
different phone sets used in our experiments. As we emphasized earlier we look at phones produced by both different and same manufacturer; hence the
difference in quantities in Table~\ref{phones}.


\begin{table}
\centering \caption{Types of phones used} \resizebox{6cm}{!}{
\begin{tabular}{|c|c|c|}
\hline
Maker&Model&Quantity\\
\hline
Apple&iPhone 5&1\\
\hline
Google&Nexus 4G&1\\
\hline
Samsung&Galaxy Note 2&1\\
\hline
Motorola&Droid A855&15\\
\hline
Sony Ericsson&W518&1\\
\hline
\end{tabular}}
\label{phones}
\end{table}

We also investigate different genres of audio excerpts. Table~\ref{audioclips} describes the different types of audio excerpts used in our experiments.
Duration of the audio clips varies from  3 to 10 seconds. The default sampling frequency of all audio excerpts is 44.1kHz unless explicitly stated otherwise. All
audio clips are stored in WAV format using 16-bit pulse-code-modulation (PCM) technique.

\begin{table}
\centering \caption{Types of audio excerpts} \resizebox{8cm}{!}{
\begin{tabular}{|c|c|}
\hline
Type&Description\\
\hline
Instrumental&Musical instruments playing together, e.g., ringtone\\
\hline
Human speech&Small segments of human speech\\
\hline
Song&Combination of human voice \& instrumental sound\\
\hline
\end{tabular}}
\label{audioclips}
\end{table}

For analysis we leverage the following audio tools and analytic modules: \emph{MIRtollbox}~\cite{mirtoolbox}, \emph{Netlab}~\cite{netlab},
\emph{Audacity}~\cite{audacity} and the Android app \emph{Hertz}~\cite{hertz}. Both MIRtoolbox and Netlab are MATLAB modules providing a rich set of
functions for analyzing and extracting audio features. Audacity and Hertz are mainly used for recording audio clips on computers and smartphones
respectively.

For analyzing and matching fingerprints we use a desktop machine with the following configuration: Intel i7-2600 3.4GHz processor with 12GiB RAM. We
found that the average time required to match a new fingerprint was around 5--10\,ms for $k$-NN classifier and around 0.5--1\,ms for GMM classifier.

\paragraphb{Evaluation metrics:}
We use standard multi-class classification metrics---\emph{precision}, \emph{recall}, and \emph{F1-score}~\cite{Sokolova2009427}---in our evaluation.
Assuming there are fingerprints from $n$ classes (i.e., $n$ distinct phones), we first compute the true positive ($TP$) rate for each class, i.e., the number of traces from the class that are classified correctly.
Similarly, we compute the false positive ($FP$) and false negative ($FN$), as the number
of wrongly accepted and wrongly rejected traces, respectively, for each class $i$ ($1\leq i\leq n$). We then compute precision, recall, and the F1-score for each
class using the following
equations:\nolinebreak
\begin{eqnarray}
\mbox{Precision,  } Pr_i &=& \frac{TP_i}{TP_i+FP_i}\\
\mbox{Recall,  } Re_i &=& \frac{TP_i}{TP_i+FN_i}\\
\mbox{F1-Score,  } \mathit{F1}_i &=& \frac{2\times Pr_i\times Re_i}{Pr_i+Re_i}
\end{eqnarray}

The F1-score is the harmonic mean of precision and recall; it provides a good measure of overall classification performance, since precision and recall represent a tradeoff: a more conservative classifier that rejects more instances will have higher precision but lower recall, and vice-versa. To obtain the overall
performance of the system we compute average values using the following equations:\nolinebreak
\begin{eqnarray}
\mbox{Avg. Precision,  } \mathit{AvgPr} &=& \frac{\sum_{i=1}^{n}Pr_i}{n}\\
\mbox{Avg. Recall,  } \mathit{AvgRe} &=& \frac{\sum_{i=1}^{n}Re_i}{n}\\
\mbox{Avg. F1-Score,  } \mathit{AvgF1} &=& \frac{2\times AvgPr\times AvgRe}{AvgPr+AvgRe}
\end{eqnarray}

Each audio excerpt is recorded/played 10 times, 50\% of which is used for training and the remaining 50\% is used for testing. We report the maximum
evaluations obtained by varying the number of neighbors ($k$) from 1 to 5 for the $k$-NN classifier and considering 1 to 5 Gaussian distributions per
class. Since GMM parameters are produced by the randomized EM algorithm, we
perform 10 parameter-generation runs for each instance and report the average classification performance.\footnote{We also computed the
95\% confidence interval, but we found it to be less than 0.01.}

\subsection{Process of Fingerprinting Speakers}\label{testspeaker}

\begin{figure}[!htb]
\centering
\includegraphics[width=1.0\columnwidth]{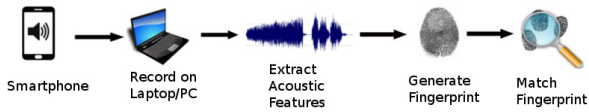}
\caption{Steps of fingerprinting speakers.} \label{speakerfingerprint}
\end{figure}

An attacker can leverage our algorithms to passively observe audio emitted from device speakers (e.g., ringtones), in public environments. To
investigate this, we first look at fingerprinting speakers integrated inside smartphones. For fingerprinting speakers we record audio clips played
from
smartphones onto a laptop and we then extract acoustic features from the recorded audio excerpts as shown in Figure~\ref{speakerfingerprint}. We look
at both devices manufactured by different vendors and the same vendor.

\subsection{Process of Fingerprinting Microphones}\label{testmic}

\begin{figure}[!htb]
\centering
\includegraphics[width=1.0\columnwidth]{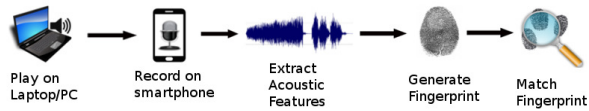}
\caption{Steps of fingerprinting microphones.} \label{microphonefingerprint}
\end{figure}

Attackers may also attempt to fingerprint devices by observing imperfections in device microphones, for example by convincing the user to install an
application on their phone, which can observe inputs from the device's microphone. To investigate feasibility of this attack, we will next look at
fingerprinting microphones embedded in smartphones. To do this, we record audio clips played from a laptop onto smartphones as shown in
Figure~\ref{microphonefingerprint}. Again we look at both devices manufactured by different vendors and the same
vendor.

\subsection{Fingerprinting Devices From Different Vendors}{\label{diffvendor}}
In this section we look at fingerprinting smartphones manufactured by five different vendor. We look at fingerprinting the devices through both
microphone and speaker.

\subsubsection{Fingerprinting Speaker}
\begin{figure}[!h]
\centering
\begin{tabular}{c}
\includegraphics[width=1.0\columnwidth]{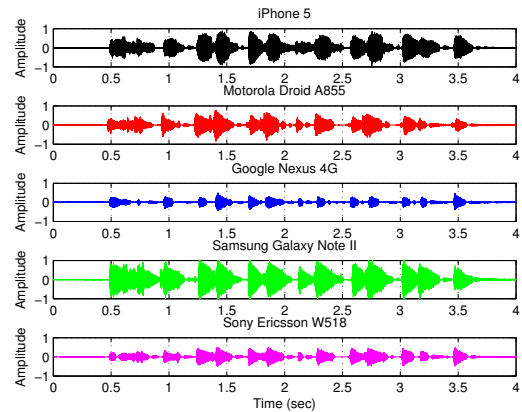}\\
(a)\\
\includegraphics[width=0.9\columnwidth]{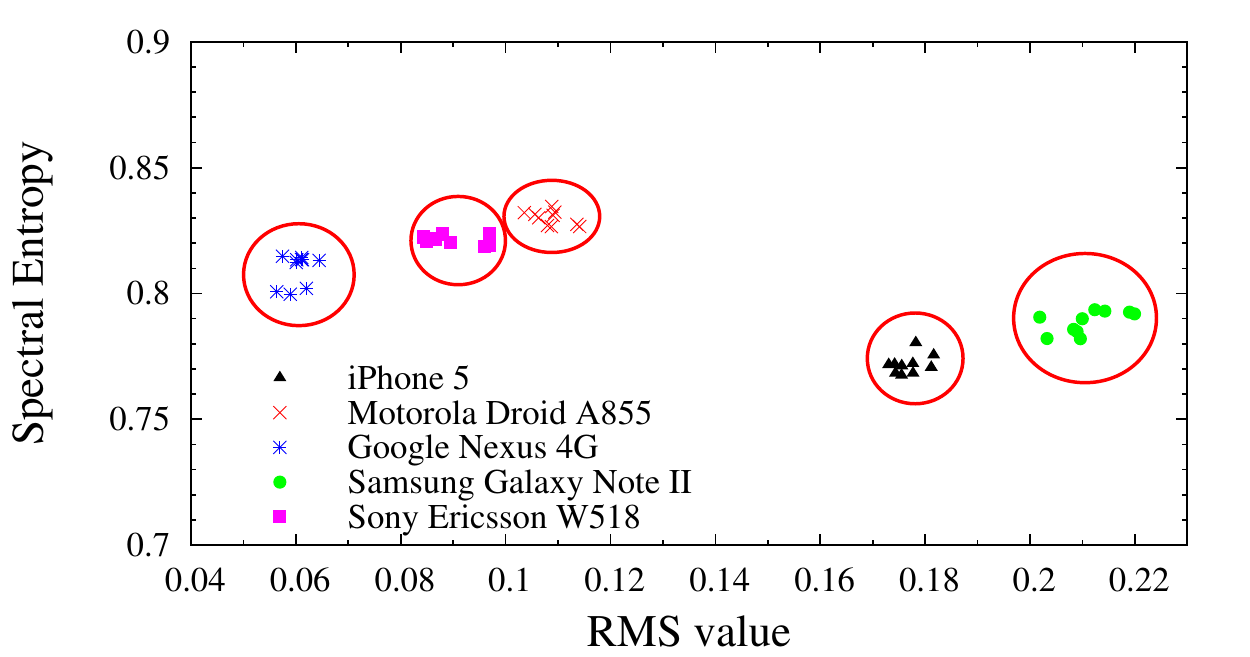}\\
(b)
\end{tabular}
\caption{a) Audio sample taken from five different handsets. b) Plotting audio samples taken from five different handsets using acoustic features ---
signal RMS value and spectral entropy.} \label{diffvendors}
\end{figure}

{\label{diffspeaker}}
We found fingerprinting smartphones manufactured by different vendors is relatively simpler compared to fingerprinting devices manufactured by the
same vendor. The main reason behind this is that the sensitivity of the speaker volume of different smartphones were quite different making it easier
to track them. Figure~\ref{diffvendors}(a) shows an audio sample played from five different smartphones. As we see the signal strength of the audio
signals are quite different from each other. Hence, simple acoustic features like RMS value and spectral entropy are good enough to obtain good
clusters of data points. Figure~\ref{diffvendors}(b) shows a plot of spectral entropy vs.\ RMS value for 50 samples of an audio excerpt (10 samples
from each handset). We see that acoustic features like spectral entropy and RMS value generate good clusters for each type of smartphone.

We test our fingerprinting approach using three different types of audio excerpts. Each audio sample is recorded 10 times giving us a total of 50
samples from the five handsets. 50\% of the samples are used for training and the remaining 50\% are used for testing, and we repeat this procedure
for the three different types of audio excerpt. Table~\ref{diffset-speaker} summarizes our findings (values are reported as percentages). We simply
use signal RMS value and spectral entropy as input features for the $k$-NN classifier, while for the GMM classifier we added MFCCs as an additional
feature because doing so increased the GMM classifier's success rate. From Table~\ref{diffset-speaker} we see that we can successfully (with a
precision rate of 100\%) identify which audio clip came from which smartphone. Thus fingerprinting smartphones manufactured by different vendors seems
very much feasible using only 2 to 3 acoustic features.

\begin{table}[h]
\centering \caption{Fingerprinting different smartphones using speaker output} \resizebox{8cm}{!}{
\begin{threeparttable}
\begin{tabular}{|c|c|c|c|c|c|c|}
\hline
\multirow{2}{*}{Audio}&\multicolumn{3}{c|}{$k$-NN}&\multicolumn{3}{c|}{GMM}\\
\cline{2-7}
\multirow{2}{*}{Type}&\multicolumn{3}{c|}{Features [1,5]\tnote{$\ast$}}&\multicolumn{3}{c|}{Features [1,5,13]\tnote{$\ast$}}\\
\cline{2-7}
&$AvgPr$&$AvgRe$&$AvgF1$&$AvgPr$&$AvgRe$&$AvgF1$\\
\hline
Instrumental&100&100&100&100&100&100\\
\hline
Human speech&100&100&100&100&100&100\\
\hline
Song&100&100&100&100&100&100\\
\hline
\end{tabular}
{\footnotesize $\ast$ Feature numbers taken from Table~\ref{allfeatures}}
\end{threeparttable}}
\label{diffset-speaker}
\end{table}

\subsubsection{Fingerprinting Microphone}
Similar to speakers, we find microphone properties differ quite substantially across vendors, simplifying fingerprinting. In particular, the
sensitivity of the microphones of the five handsets were different. As a result, when the same audio clip is recorded on the phones their respective
RMS value and spectral entropy are distinguishably different,  making it possible to fingerprint smartphones through microphones. To test our
hypothesis we again test our fingerprinting approach using three different types of audio excerpts. Each audio sample is recorded 10 times, 50\% of
which are used for training and the remaining 50\% are used for testing. Table~\ref{diffset-microphone} summarizes our findings (values are reported
as percentages). We use the same set of features as we did in section {\ref{diffspeaker}} and we see similar outcomes. These results suggest that
smartphones can be successfully fingerprinted through microphones.

\begin{table}
\centering \caption{Fingerprinting different smartphones using mic} \resizebox{8cm}{!}{
\begin{threeparttable}
\begin{tabular}{|c|c|c|c|c|c|c|}
\hline
\multirow{2}{*}{Audio}&\multicolumn{3}{c|}{$k$-NN}&\multicolumn{3}{c|}{GMM}\\
\cline{2-7}
\multirow{2}{*}{Type}&\multicolumn{3}{c|}{Features [1,5]\tnote{$\ast$}}&\multicolumn{3}{c|}{Features [1,5,13]\tnote{$\ast$}}\\
\cline{2-7}
&$AvgPr$&$AvgRe$&$AvgF1$&$AvgPr$&$AvgRe$&$AvgF1$\\
\hline
Instrumental&96.7&96&96.3&96.7&96&96.3\\
\hline
Human speech&93.3&92&92.6&96.7&96&96.3\\
\hline
Song&96.7&96&96.3&100&100&100\\
\hline
\end{tabular}
{\footnotesize $\ast$ Feature numbers taken from Table~\ref{allfeatures}}
\end{threeparttable}}
\label{diffset-microphone}
\end{table}

\subsection{Fingerprinting Devices of the Same Model}{\label{samevendor}}
In this section we look at fingerprinting smartphones manufactured by the same vendor. We found that this was relatively a tougher problem and as such we first explore all the 15 acoustic features listed in Table~\ref{allfeatures} to determine the
dominating subset of features. Next, we carry out our fingerprinting task using the dominant subset of acoustic features. We again fingerprint
devices through both microphone and speaker. Note that the audio excerpts used for feature exploration in Section {\ref{SFfeature}} and the ones used
for evaluating our fingerprinting approach in Section~{\ref{same-vendor-speakers}} and {\ref{same-vendor-mics}} are not identical. We use different
audio excerpts, though belonging to the same categories as listed in Table~\ref{audioclips}, so as to \emph{not bias} our evaluations.

\subsubsection{Feature Exploration}\label{SFfeature}
At first glance, it seems that we should use all features at our disposal to identify device types. However, including too many features can {\em
worsen} performance in practice, due to their varying accuracies and potentially-conflicting signatures.  Hence, in this section, we explore all the
15 audio features described in Section {\ref{audio-feature}} and identify the {\em dominating subset} of all the features, i.e., which combination of
features should be used. For this purpose we adopt a well known machine learning strategy known as \emph{feature
selection}~\cite{Guyon:2003,Yang:1997}. Feature selection is the process of reducing dimensionality of data by selecting only a subset of the
relevant features for use in model construction. The main assumption in using feature selection technique is that the data may contain redundant
features. Redundant features are those which provide no additional benefit than the currently selected features. Feature selection techniques are a
subset of the more general field of feature extraction, however, in practice they are quite different from each other. Feature extraction creates new
features as functions of the original features, whereas feature selection returns a subset of the features. Feature selection is preferable to
feature extraction when the original units and meaning of features are important and the modeling goal is to identify an influential subset. When the
features themselves have different dimensionality, and numerical transformations are inappropriate, feature selection becomes the primary means of
dimension reduction.

\begin{table*}[tbp]
\centering \caption{Feature exploration using sequential forward selection technique} \resizebox{14cm}{!}{
\begin{tabular}{|c|c|c|c|c|c|c|c|c|c|c|}
\hline
\multirow{3}{*}{\#}&\multirow{3}{*}{Feature}&\multicolumn{3}{c|}{\multirow{2}{*}{Avg. Feature-Extraction Time ($msec$)}}&\multicolumn{6}{c|}{Maximum
F1-Score (\%)}\\
\cline{6-11}
&&\multicolumn{3}{c|}{}&\multicolumn{2}{c|}{Instrumental}&\multicolumn{2}{c|}{Human Speech}&\multicolumn{2}{c|}{Song}\\
\cline{3-11}
&&Instrumental&{Human Speech}&{Song}&$k$-NN&GMM&$k$-NN&GMM&$k$-NN&GMM\\
\hline
1&RMS&9.26&10.01&11.23&21.8&17&37.9&34.4&20.1&26.2\\
\hline
2&ZCR&9.48&10.61&12.57&17.3&15.2&34.4&31.6&13&7.2\\
\hline
3&Low-Energy-Rate&29.28&32.62&39.27&9.4&39.6&18.3&13.7&21.8&19.2\\
\hline
4&Spectral Centroid&79.40&79.61&88.51&39.4&37.3&33.8&30.8&39.9&40.3\\
\hline
5&Spectral Entropy&57.54&46.58&61.88&39.6&39.6&30.4&38.7&33.9&26.1\\
\hline
6&Spectral Irregularity&6519.89&2387.04&15348.45&36&32.2&23.8&25.4&14.1&14.8\\
\hline
7&Spectral Spread&80.12&69.19&108.23&44.4&39.6&35.2&31.7&35.2&38.4\\
\hline
8&Spectral Skewness&120.29&109.26&179.33&32&41.7&30.1&34.3&31.5&40.4\\
\hline
9&Spectral Kurtosis&136.86&131.17&154.03&43&39.6&34.2&39.2&31.1&36.8\\
\hline
10&Spectral Rolloff&73.16&52.08&65.70&57.3&50.6&29&30.5&38.7&39.4\\
\hline
11&Spectral Brightness&63.91&45.51&59.94&23.5&19.9&33.5&33.5&18.5&17.9\\
\hline
12&Spectral Flatness&76.48&57.38&71.79&41.9&35.8&37.1&39.4&32.4&29.8\\
\hline
13&MFCCs&268.86&229.81&413.16&\textbf{92.4}&\textbf{97.2}&\textbf{98.8}&\textbf{98.8}&\textbf{90}&\textbf{91.4}\\
\hline
14&Chromagram&56.07&76.87&69.68&57.1&49.6&95.2&96.7&80.1&79.7\\
\hline
15&Tonal Centroid&79.54&99.95&85.79&57.1&46.1&93.7&95.2&63.6&53.7\\
\hline
\multicolumn{5}{|c|}{\multirow{2}{*}{Sequential Feature Selection}}&{[13,14]}&{[13,14]}&[13]&[13,14]&[13,7]&[13,14]\\
\multicolumn{5}{|c|}{}&96.3&97.7&98.8&100&92.6&94.1\\
\hline
\end{tabular}}
\label{rankedfeature}
\end{table*}

Feature selection involves the maximization of an objective function as it searches through the possible candidate
subsets. Since exhaustive evaluation of all possible subsets are often infeasible ($2^N$ for a total of $N$ features) different heuristics are
employed. We use a greedy search strategy known as \emph{sequential forward selection} (SFS) where we start off with an empty set and
sequentially add the features that maximize our objective function. The pseudo code of our feature selection algorithm is described in
Algorithm~\ref{feature-selection}.

\begin{algorithm}[!h]
\caption{Sequential Feature Selection}\label{feature-selection}
\begin{algorithmic}
\STATE  {\bf Input:} {Input feature set $F$}

\STATE {\bf Output:} {Dominant feature subset $D$}

\STATE{$F1\_score\leftarrow []$}
\FOR{$f\in F$}
    \STATE {$F1\_score[f] \leftarrow Classify(f)$ }
\ENDFOR

\STATE {$F'\leftarrow sort(F,F1\_score)$  \#In descending order}

\STATE {$max\_score\leftarrow 0$}
\STATE{$D \leftarrow \varnothing$}

\FOR{$f\in F'$}
    \STATE {$D \leftarrow D \cup f$ }
    \STATE {$temp \leftarrow Classify(D)$ }
    \IF{$temp>max\_score$}
        \STATE{$max\_score \leftarrow temp$}
    \ELSE
        \STATE{$D \leftarrow D-\{f\}$}
    \ENDIF
\ENDFOR

\STATE {\textbf{return} $D$}
\end{algorithmic}
\end{algorithm}

The algorithm works as follows. First, we compute the F1-score that can be achieved by each feature individually. Next, we sort the feature set based
on the achieved F1-score in descending order. Then, we iteratively add features starting from the most dominant one and compute the F1-score of the
combined feature subset. If adding a feature increases the F1-score seen so far we move on to the next feature, else we remove the feature under
inspection. Having traversed through the entire set of features we return the subset of features that maximizes our device classification task. Note
that this is a greedy approach, therefore, the generated subset might not always provide optimal F1-score. However, for our purpose, we found this
approach to perform well, as we demonstrate in latter sections.

We test our feature selection algorithm for all three types of audio excerpts listed in Table~\ref{audioclips}. We evaluate the F1-score using both
$k$-NN and GMM classifiers. Table~\ref{rankedfeature} highlights the maximum F1-score obtained by varying $k$ from 1 to 5 (for $k$-NN classifier) and
also considering 1 to 5 gaussian distributions per class (for GMM classifier). To obtain the fingerprinting data we record audio clips played from 15
Motorola Droid A855 handsets. Each type of audio is recorded 10 times giving us a total of 150 samples from the 15 handsets; 50\% of which (i.e., 5
samples per class) are used for training and the remaining 50\% are used for testing. All the training samples are labeled with their corresponding
handset identifier. Both classifiers return the class label for each audio clip in the test set and from that we compute F1-score.
Table~\ref{rankedfeature} shows the maximum F1-score achieved by each acoustic feature for the three different types of audio excerpt. We also
provide the time required to extract each feature. The table highlights the subset of features selected by our sequential feature selection algorithm
and their corresponding F1-score. We find that \emph{MFCCs} are the dominant feature for all category of audio excerpt. \emph{Chromagram} also
generates high F1-score.

\begin{figure*}[!htb]
\centering
\begin{tabular}{ccc}
\includegraphics[width=0.32\linewidth]{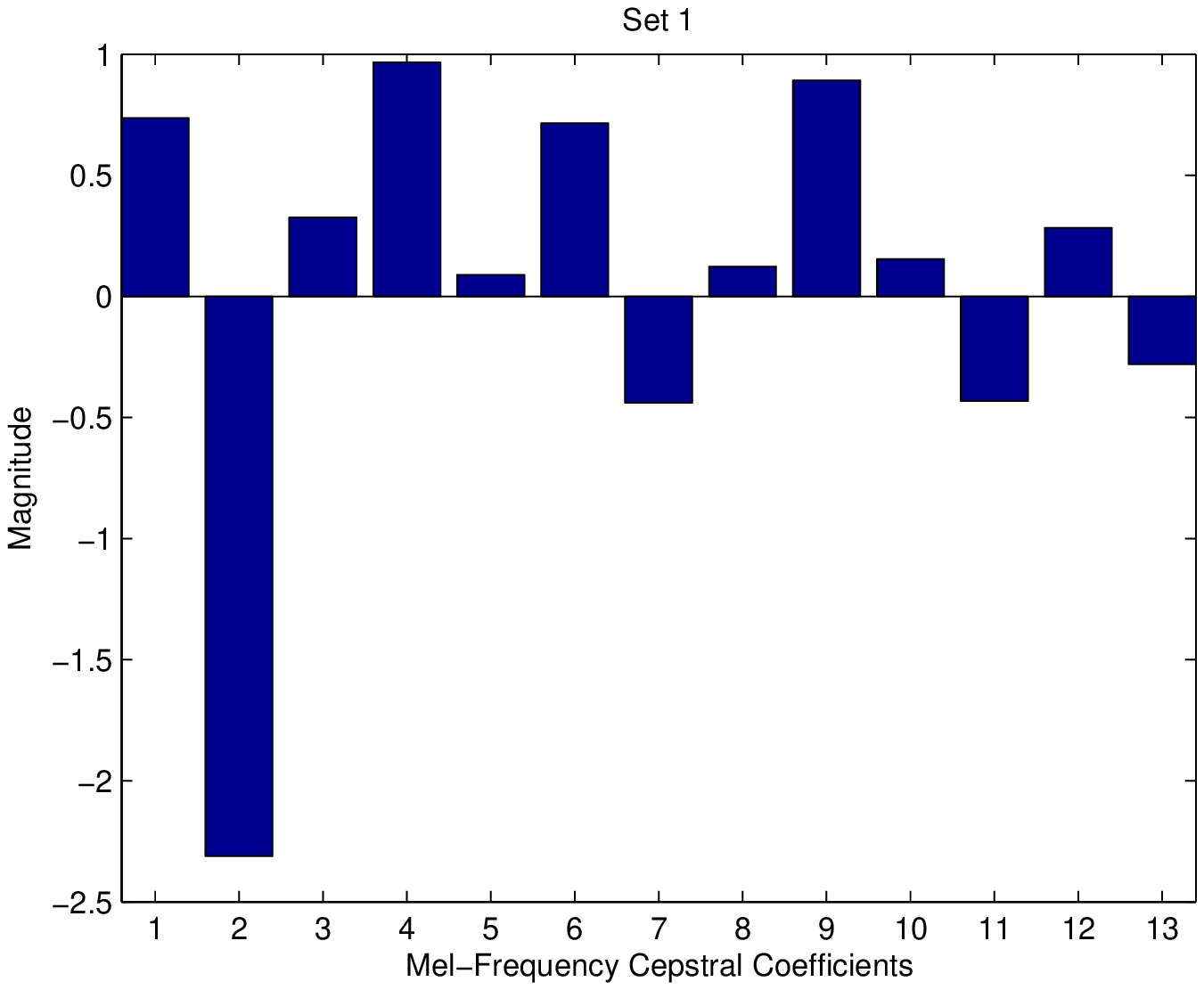}&\includegraphics[width=0.32\linewidth]{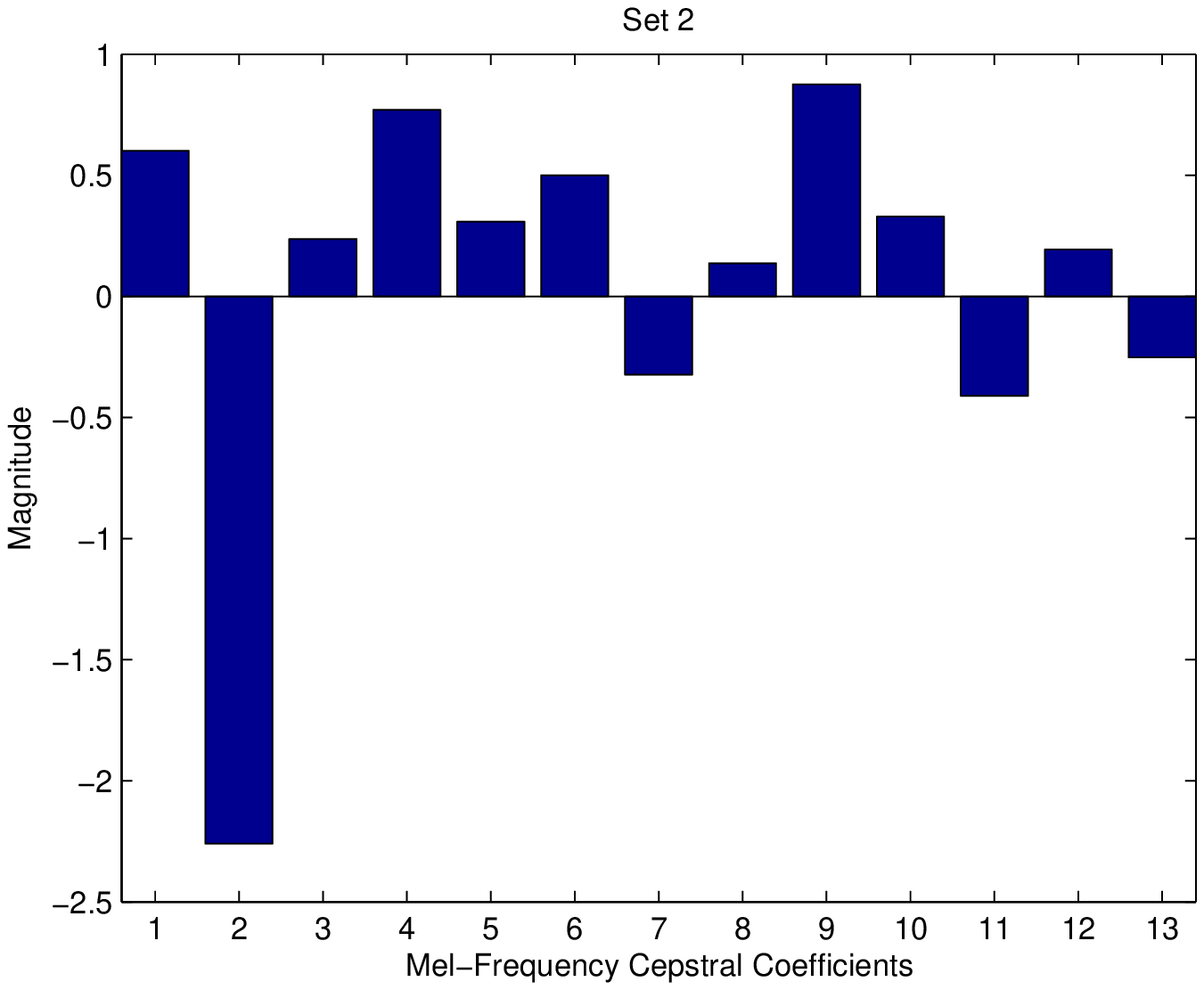}&\includegraphics[width=0.32\linewidth]{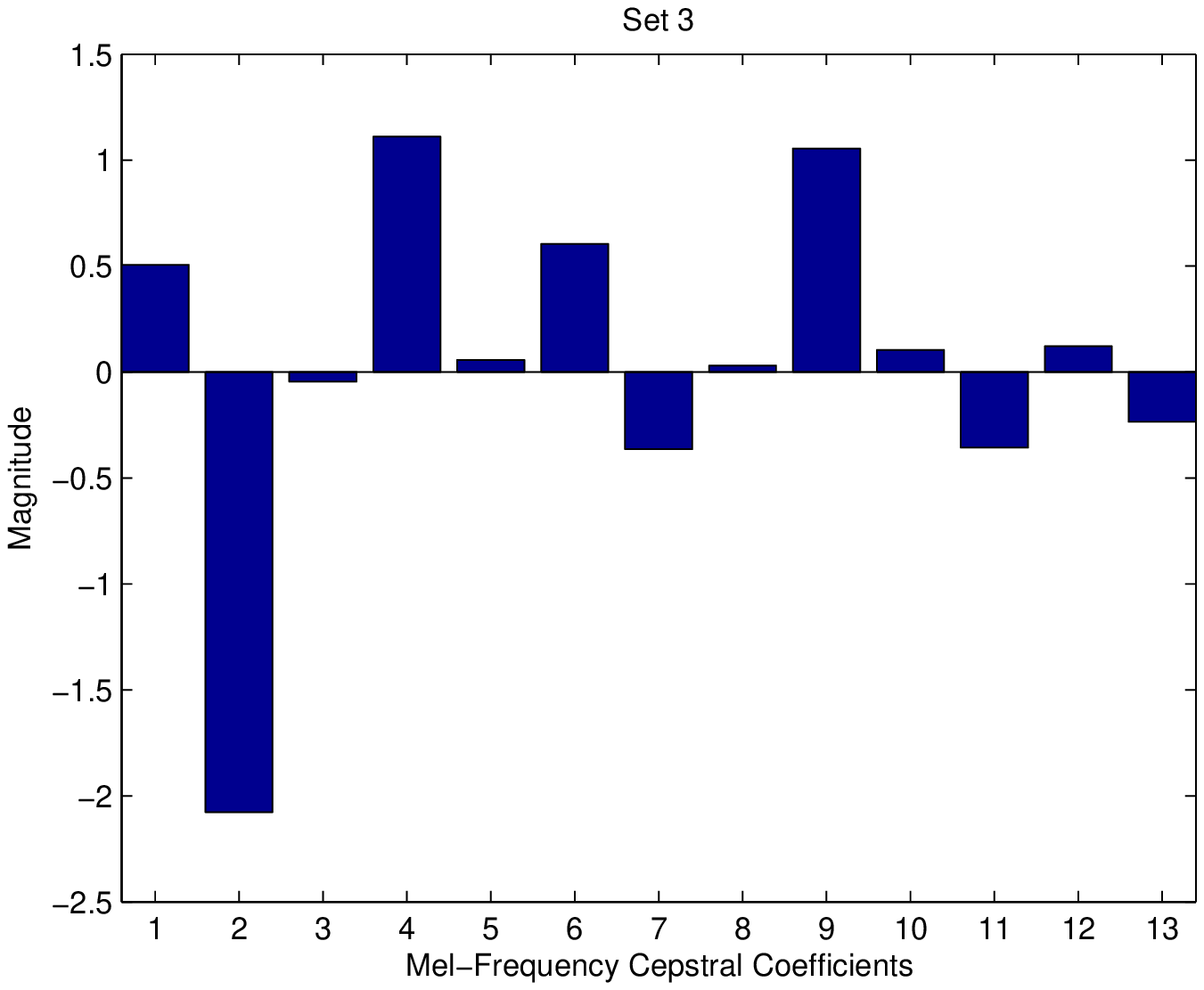}
\end{tabular}
\caption{MFCCs of the same audio sample taken from three different handsets manufactured by the same vendor. We can see that some of the coefficients
vary significantly, thus enabling us to exploit this feature to fingerprint smartphones.} \label{MFCCs}
\end{figure*}

To give a better understanding of why \emph{MFCCs} are the dominant acoustic features we plot the MFCCs of a given audio excerpt from three different
handsets on Figure~\ref{MFCCs}. All the coefficients are ranked in the same order for the three handsets. We can see that the magnitude of the
coefficients vary across the three handsets. For example coefficients 3 and 5 vary significantly across the three handsets. This makes MFCCs a
prevalent choice for fingerprinting smartphones.

\subsubsection{Fingerprinting Speakers}{\label{same-vendor-speakers}}

We now look at fingerprinting smartphones manufactured by the same vendor. For these set of experiments we use 15 Motorola Droid A855 handsets.
Table~\ref{sameset-speaker} highlights our findings. We again test our fingerprinting approach against three different forms of audio excerpt. We use
the acoustic features obtained from our sequential feature selection algorithm as listed in Table~\ref{rankedfeature}. From
Table~\ref{sameset-speaker}, we see that we can achieve an F1-score of over 94\% in identifying which audio clip originated from which handset. Thus
fingerprinting smartphones through speaker seems to be a viable option.

\begin{table}[h]
\centering \caption{Fingerprinting similar smartphones using speaker output} \resizebox{8cm}{!}{
\begin{threeparttable}
\begin{tabular}{|c|c|c|c|c|c|c|c|c|}
\hline
{Audio}&\multicolumn{4}{c|}{$k$-NN}&\multicolumn{4}{c|}{GMM}\\
\cline{2-9}
{Type}&Features\tnote{$\ast$}&$AvgPr$&$AvgRe$&$AvgF1$&Features\tnote{$\ast$}&$AvgPr$&$AvgRe$&$AvgF1$\\
\hline
Instrumental&[13,14]&96.7&96&96.3&[13,14]&98.3&98&98.1\\
\hline
Human speech&[13]&98.9&98.7&98.8&[13,14]&98.9&98.7&98.8\\
\hline
Song&[13,7]&93.7&92&92.8&[13,14]&95.6&93.3&94.4\\
\hline
\end{tabular}
{\footnotesize $\ast$ Feature numbers taken from Table~\ref{rankedfeature}}
\end{threeparttable}}
\label{sameset-speaker}
\end{table}

\subsubsection{Fingerprinting Microphone}{\label{same-vendor-mics}}
We now investigate fingerprinting smartphones made by the same vendor through microphone-sourced input. We use 15 Motorola Droid A855 handsets for
these experiments. We use the features obtained through Algorithm~\ref{feature-selection} which are listed in Table~\ref{rankedfeature}.
Table~\ref{sameset-micrphone} summarizes our findings. We see similar results compared to fingerprinting speakers. We were able to achieve an
F1-score of 93\% in identifying the handset from which the audio excerpt originated. Thus fingerprinting smartphones through microphones also appears
to be feasible.

\begin{table}[h]
\centering \caption{Fingerprinting similar smartphones using microphone} \resizebox{8cm}{!}{
\begin{threeparttable}
\begin{tabular}{|c|c|c|c|c|c|c|c|c|}
\hline
{Audio}&\multicolumn{4}{c|}{$k$-NN}&\multicolumn{4}{c|}{GMM}\\
\cline{2-9}
{Type}&Features\tnote{$\ast$}&$AvgPr$&$AvgRe$&$AvgF1$&Features\tnote{$\ast$}&$AvgPr$&$AvgRe$&$AvgF1$\\
\hline
Instrumental&[13,14]&93.7&92&92.8&[13,14]&94.1&92&93\\
\hline
Human speech&[13]&98.9&98.7&98.8&[13,14]&98.9&98.7&98.8\\
\hline
Song&[13,7]&93.9&93.3&93.6&[13,14]&96.1&95.2&95.6\\
\hline
\end{tabular}
{\footnotesize $\ast$ Feature numbers taken from Table~\ref{rankedfeature}}
\end{threeparttable}}
\label{sameset-micrphone}
\end{table}

\subsection{Sensitivity Analysis}\label{sensitivity}

In this section we investigate how different factors such as audio sampling rate, training set size, the distance from audio source to recorder, and
background noise impact our fingerprinting performance. Such investigations will help us determine the conditions under which our fingerprinting
approach will be feasible. For the following set of experiments we will only focus on fingerprinting smartphones from the same vendor and we only
consider fingerprinting speakers as we saw almost identical outcomes for fingerprinting microphones. We also consider recording only ringtones (i.e.,
an audio clip belonging to our defined Instrumental category) for the following experiments. Since we are recording ringtones we only use the
features highlighted in Table~\ref{rankedfeature} under `Instrumental' category.

\subsubsection{Impact of Sampling Rate}
\begin{figure}[h]
\centering
\includegraphics[width=0.80\columnwidth]{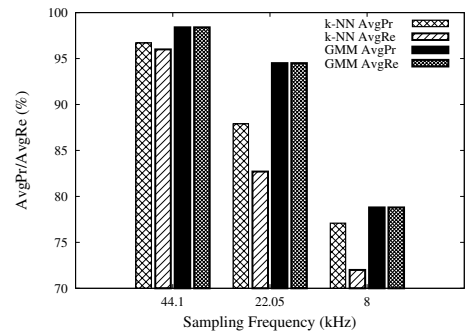}
\caption{Impact of sampling frequency on precision/recall.} \label{diffrate}
\end{figure}

First, we investigate how the sampling rate of audio signals impacts our fingerprinting precision. To do this, we record a ringtone at the following
three frequencies: 8kHz, 22.05kHz and 44.1kHz. Each sample is recorded 10 times with half of them being used for training and the other half for
testing. Figure~\ref{diffrate} shows the average precision and recall obtained under different sampling rates. As we can see from the figure, as
sampling frequency decreases, the precision/recall also goes down. This is understandable, because the higher the sampling frequency the more
fine-tuned information we have about the audio sample. However, the default sampling frequency on most handheld devices today is
44.1kHz~\cite{smartphonespeakers}, with some of the latest models adopting even higher sampling rates~\cite{dacs}. We, therefore, believe sampling
rate will not impose an obstacle to our fingerprinting approach, and in future we will be able to capture more fine grained variations with the use
of higher sampling rates.

\subsubsection{Varying Training Size}
Next, we consider performance of the classifiers in the presence of limited training data. For this experiment we vary the training set size from
10\% to 50\% (i.e., from 1 to 5 samples per class) of all available samples. Table~\ref{difftrainsize} shows the evolution of the F1-score as
training set size is increased (values are reported as percentages). We see that as the training set size increases the F1-score also rises which is
expected. However, we see that with only three samples per class we can achieve an F1-score of over 90\%. This suggests that we do not need too many
training samples to construct a good predictive model.


\begin{table}[h]
\centering \caption{Impact of varying training size} \resizebox{7.5cm}{!}{
\begin{threeparttable}
\begin{tabular}{|c|c|c|c|c|c|c|}
\hline
{Training}&\multicolumn{3}{c|}{$k$-NN}&\multicolumn{3}{c|}{GMM}\\
\cline{2-7}
{samples}&\multicolumn{3}{c|}{Features [13,14]\tnote{$\ast$}}&\multicolumn{3}{c|}{Features [13,14]\tnote{$\ast$}}\\
\cline{2-7}
per class&$AvgPr$&$AvgRe$&$AvgF1$&$AvgPr$&$AvgRe$&$AvgF1$\\
\hline
1&42&49.3&45.3&50&53.3&51.6\\
\hline
2&79.2&80&79.6&80.4&80&80.2\\
\hline
3&91.3&89.3&90.2&91.7&89.3&90.5\\
\hline
4&95.3&94.7&95&95.6&94.7&95.1\\
\hline
5&96.7&96&96.3&98.3&98&98.1\\
\hline
\end{tabular}
{\footnotesize $\ast$ Feature numbers taken from Table~\ref{rankedfeature}}
\end{threeparttable}}
\label{difftrainsize}
\end{table}

\subsubsection{Varying Distance between Audio Source and Recorder}

\begin{figure}[h]
\centering
\includegraphics[width=0.6\columnwidth]{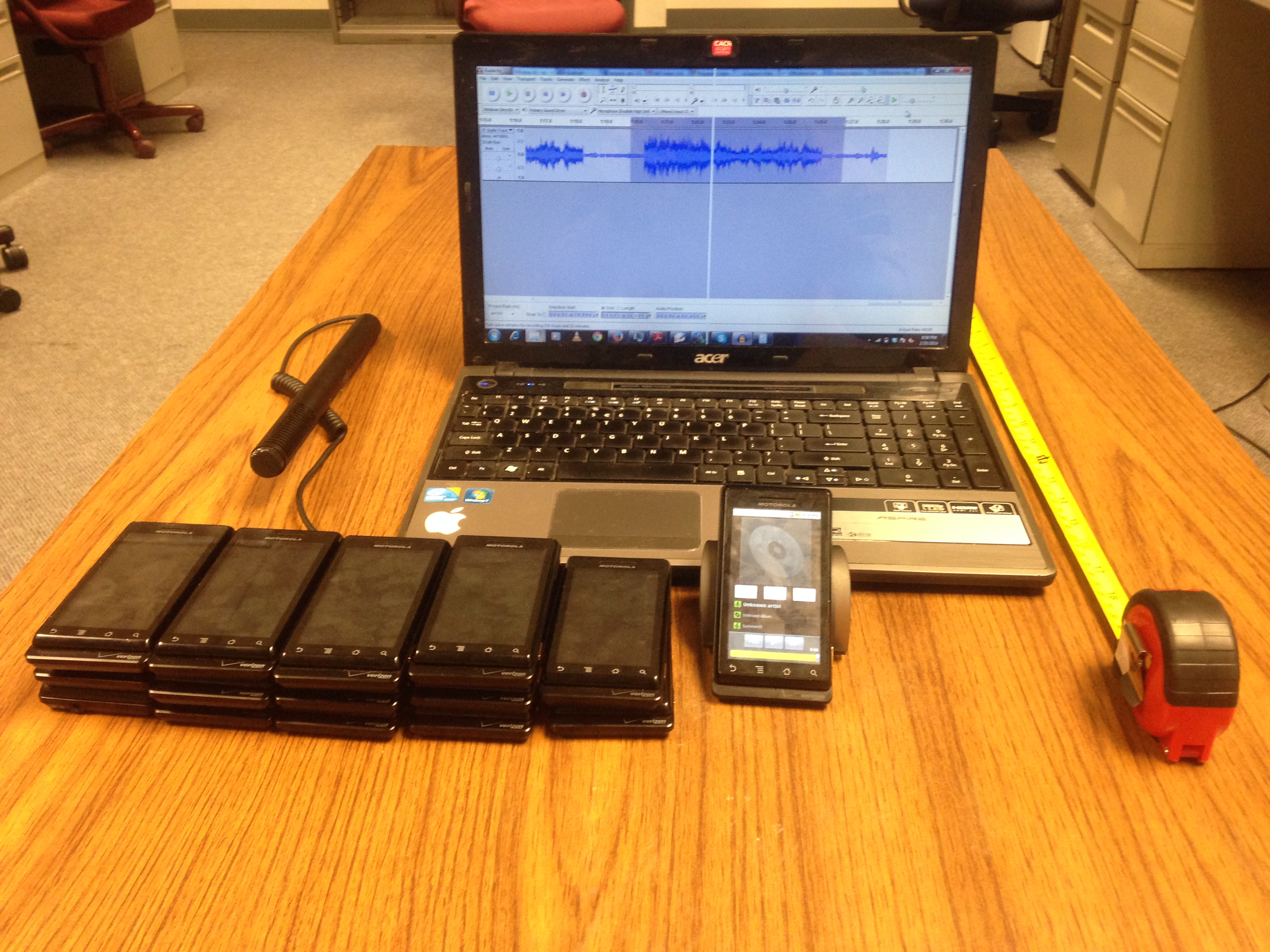}
\caption{Experimental setup for varying the distance between the smartphone and microphone.}
\label{distance}
\end{figure}

Next, we inspect the impact of distance between the audio source (i.e., smartphone) and recorder (i.e., laptop/PC) on fingerprinting
precision/recall. For this experiment we use a separate external microphone as the signal capturing capacity of microphones embedded inside laptops
degrades drastically as distance increases. We use the relatively inexpensive (\$44.79) Audio-Technica ATR-6550 shotgun microphone for this
experiment and vary the distance between the external microphone and smartphone from 0.1 meter to 5 meters. Figure~\ref{distance} shows the
experimental setup and Table~\ref{distancevary} summarizes the F1-scores obtained as the distance between the smartphone and microphone varies. We
see that as distance increases, F1-score decreases. This is expected, because as the distance between the smartphone and microphone increases, the
harder it becomes to capture the minuscule deviations between audio samples. However, we see that even up to two meters distance we can achieve an
F1-score of 93\%. This suggests that our device fingerprinting approach works only up to a certain distance using any commercial microphones.
However, using specialized microphones, such as parabolic microphones (usually used in capturing animal sounds from a far distance) could help in
increasing the fingerprinting precision at even longer distances.

\begin{table}[h]
\centering \caption{Impact of varying distance} \resizebox{7.5cm}{!}{
\begin{threeparttable}
\begin{tabular}{|c|c|c|c|c|c|c|}
\hline
\multirow{2}{*}{Dintance}&\multicolumn{3}{c|}{$k$-NN}&\multicolumn{3}{c|}{GMM}\\
\cline{2-7}
\multirow{2}{*}{(meters)}&\multicolumn{3}{c|}{Features [13,14]\tnote{$\ast$}}&\multicolumn{3}{c|}{Features [13,14]\tnote{$\ast$}}\\
\cline{2-7}
&$AvgPr$&$AvgRe$&$AvgF1$&$AvgPr$&$AvgRe$&$AvgF1$\\
\hline
0.1&96.7&96&96.3&98.3&98&98.1\\
\hline
1&92.7&91.5&92&95.2&94.7&94.9\\
\hline
2&88.2&87.6&87.9&94.5&92&93.2\\
\hline
3&76.7&76&76.3&78.9&84&81.4\\
\hline
4&70.2&64&67&76.8&76&76.4\\
\hline
5&64.5&62.7&63.6&77&73.3&75.1\\
\hline
\end{tabular}
{\footnotesize $\ast$ Feature numbers taken from Table~\ref{rankedfeature}}
\end{threeparttable}}
\label{distancevary}
\end{table}

\subsubsection{Impact of Ambient Background Noise}

In this section we investigate how ambient background noise impacts the performance of our fingerprinting technique. For this experiment we consider
scenario types were there is a crowd of people using their smart devices and we are trying to fingerprint those devices by capturing audio signals
(in this case ringtones) from the surrounding environment. Table~\ref{background} highlights the four different scenarios that we are considering. To
capture audio signals under such scenarios -- external speakers (2 pieces), placed between the smartphone and microphone, were constantly replaying
the respective ambient noise in the background while recording of ringtones played from different handsets were taking place. We consider a distance
of two meters from the audio source to recorder. The ambient background sounds were obtained from PacDV~\cite{sounds} and SoundJay~\cite{soundjay}.
Table~\ref{background} shows our findings (values are reported as percentages). We can see that even in the presence of various background noise we
can achieve an F1-score of over 91\%.

\begin{table}[h]
\centering \caption{Impact of ambient background noise} \resizebox{8cm}{!}{
\begin{threeparttable}
\begin{tabular}{|c|c|c|c|c|c|c|}
\hline
\multirow{3}{*}{Environments}&\multicolumn{3}{c|}{$k$-NN}&\multicolumn{3}{c|}{GMM}\\
\cline{2-7}
&\multicolumn{3}{c|}{Features [13,14]\tnote{$\ast$}}&\multicolumn{3}{c|}{Features [13,14]\tnote{$\ast$}}\\
\cline{2-7}
&$AvgPr$&$AvgRe$&$AvgF1$&$AvgPr$&$AvgRe$&$AvgF1$\\
\hline
Shopping Mall&88.8&85.3&87&95.1&93.3&94.2\\
\hline
Restaurant/Cafe&90.5&89.7&90.1&92.5&90.7&91.6\\
\hline
City Park&91.7&90&90.8&95.2&94.1&94.6\\
\hline
Airport Gate&91.3&89.5&90.4&94.5&93.3&93.9\\
\hline
\end{tabular}
{\footnotesize $\ast$ Feature numbers taken from Table~\ref{rankedfeature}}
\end{threeparttable}}
\label{background}
\end{table}

\section{Applications}{\label{applications}}

Fingerprinting smart devices can be thought of as a double-edged sword when it comes to device security. On one hand, it can jeopardize privacy, as
it allows remote identification without user awareness. On the other hand, it could potentially be used to enhance authentication of physical
devices. We discuss these potential applications below.


\subsection{Multi-factor Authentication}
Conventional computing systems authenticate users by verifying some static factors such as user generated passwords (which may be coupled with
additional security questions like pin code or phone number). A password consists of a string of characters, remembered by the human user, which can
be provided as a proof of identity. However, passwords are vulnerable to guessing algorithms. Moreover, if passwords ever leak they potentially open
an opportunity for an unauthenticated user to get access to the system. Often systems do not incorporate mechanisms to verify whether the
authenticated user is using an authorized {\em device}. Modern highly-secure organizations (e.g., military and department of defense) are therefore,
moving towards using various forms of active authentication for their employees~\cite{darpa}.

Device fingerprinting can be used to provide a multi-factor authentication framework that will enable a system administrator to validate whether
authenticated users are using their allocated devices to log in into the system. This scenario of course is applicable to high-security conscious
organizations where tracking authenticated users is not against any privacy violation. This can be done by leveraging our techniques, for example by
instructing the user's device to record an audio sample broadcast over the PA system, or transmit an audio session over the phone. Alternatively, the
device may be able to ``fingerprint itself'', by playing a received small audio clip out its speaker, simultaneously recording via its microphone,
and then transmitting the result over the  network to the authentication server for verification\footnote{We are assuming that the user is not using
a headphone at the start of authentication.}.
In this way we can tie both user and device identity together to form a multi-factor authentication framework. As a side note this only provides
additional assurance, rather than a foolproof authentication method. However, we believe our approach is more robust than existing software based
two-factor authentication systems (e.g., for systems where you need to submit a token along with a password, if the attacker gets hold off the secret
key then he/she can generate the desired token) as it is harder to mimic hardware level imperfections.

\subsection{Device Tracking}
By the same token, an attacker can violate user privacy by via a similar approach, or installing a malicious application on the user's device, or
recording broadcasted audio in public environments.
For example, a malicious application (e.g., a game) can play small audio segments, record them using the device's microphone, and upload recorded
clips to the attacker. To do this, the application would require access to both microphone and network access
permission, but this might not be a big assumption to make: most users are unaware of the security risks associated with mobile apps and a significant portion of the users cannot
fully comprehend the full extension of all the permissions~\cite{apppermissions,Felt:2012}.

Alternatively, the attacker may sit in public environments (cafe, shopping mall), and
record broadcasted audio (speakerphone conversations, ringtones) with the intent to track and identify users.



\section{Related Work}{\label{related-work}}

Fingerprints have long been used as one of the most common biometrics in identifying human beings~\cite{Ross20032115,cole2009suspect}. The same
concept was extended to identifying and tracking unique mobile transmitters by the US government during 1960s~\cite{Langley93}. Later on with the
emergence of cellular networks people were able to uniquely identify transmitters by analyzing the externally observable characteristics of the
emitted radio signals~\cite{Riezenman2000}.

Physical devices are usually different at either the software or hardware level even if they are produced by the same vendor. In terms of software
based fingerprinting researchers have looked at fingerprinting techniques that differentiates between unique devices over a Wireless Local Area
Network (WLAN) simply through a timing analysis of 802.11 probe request frames~\cite{Desmond:2008}. Others have looked at exploiting the difference
in firmware and device driver running on IEEE 802.11 compliant devices~\cite{Franklin:2006}. 802.11 MAC headers have also been used to track unique
devices~\cite{Guo05}. Pang et al.~\cite{Pang:2007} were able to exploit traffic patterns to carry out device fingerprinting. Open source toolkits
like Nmap~\cite{nmap} and Xprobe~\cite{xprobe} can remotely fingerprint an operating system by identifying unique responses from the TCP/IP
networking stack.

Another angle to software based fingerprinting is to exploit applications like browsers to carry out device fingerprinting~\cite{Eckersley:2010}. Yen
et al.~\cite{Yen12} were successful at tracking users with high precision by analyzing month-long logs of Bing and Hotmail. Researchers have also
been able to exploit JavaScript and popular third-party plugins like Flash player to obtain the list of fonts installed in a device which then
enabled them to uniquely track users~\cite{Acar:2013}. Other researchers have proposed the use of performance benchmarks for differentiating between
JavaScript engines~\cite{MBYS11}. Furthermore, browsing history can be exploited to fingerprint and track web users~\cite{olejnik:hal-00747841}. The
downside of software based fingerprints is that such fingerprints are generated from the current configuration of the system which is not static,
rather it is likely to change over time.

Hardware based fingerprinting approaches rely on some static source of idiosyncrasies. It has been shown that network devices tends to have constant
clock skews~\cite{Moon99} and researchers have been able to exploit these clock skews to distinguish devices through TCP and ICMP
timestamps~\cite{Kohno:2005}. However, clock skew rate is highly dependent on the experimental environment. Researchers have also extensively looked
at fingerprinting the unique transient characteristics of radio transmitters (also known as Radio Frequency (RF) fingerprinting). RF fingerprinting
has been shown as a means of enhancing wireless authentication~\cite{Li:2006,Nam2011,Ureten2007,Bonne2007}. It has also been used for location
detection~\cite{Patwari:2007}. Manufacturing imperfections in network interface cards (NICs) have also been studied by analyzing analog signals
transmitted from them~\cite{Brik:2008,Gerdes06}. More recently Dey et al. have studied manufacturing idiosyncrasies inside smartphone accelerometer
to distinguish devices~\cite{accelprint}. However, their approach requires some form of external stimulation/vibration to successfully capture the
manufacturing imperfection of the on-board accelerometer. Moreover, there are different contexts in which audio prints can be more useful, e.g.,
software that is not allowed to access the accelerometer, as well as an external adversary who fingerprints nearby phones with a microphone.

Our work is inspired by the aforementioned hardware based fingerprinting works, but instead of focusing on wireless transmitters or on-board sensors
that require external stimulation, we focus on fingerprinting on-broad acoustic components like microphones and speakers.

Audio fingerprinting has a rich history of notable research works~\cite{Cano:2005}. There are studies that have looked at classifying audio excerpts
based on their content~\cite{Tzanetakis2002,Guo2003}. Others have looked at distinguishing human speakers from audio
segments~\cite{Campbell97,Bimbot2004}. There has also been work on exploring various acoustic features for audio classification~\cite{Mckinney03}.
One of the more popular applications of audio fingerprinting has been music genre and artist recognition~\cite{Li:2003,Haitsma02ahighly}.

Our work takes advantage of the large set of acoustic features that have been explored by the aforementioned works. However, instead of classifying
the content of audio segments, we are utilizing the acoustics features to capture the manufacturing imperfections of microphones and speakers
embedded in smart devices.

\section{Discussion and Limitations}{\label{limitations}}

Our approach has several limitations. First, we experimented with 15 devices manufactured by the same vendor; it is possible that a larger target
device pool would result in lower accuracy. That said, distinctions across different device types are more clear; additionally, audio fingerprints
may be used in tandem with other techniques, such as accelerometer fingerprinting~\cite{accelprint}, to better discriminate between devices.
Secondly, we evaluated our fingerprinting precision/recall under only two types of classifiers (GMM and $k$-NN). Other forms of classification such
as ensemble based approaches could possibly achieve better results, as ensemble based methods use multiple models to obtain better predictive
performance than any single classifier~\cite{ensemble}. However, as a first step we were able to achieve over 93\% precision using simple $k$-NN and
GMM classifiers, and our results may point to the concern that relatively simple techniques have a high success rate. Lastly, all the phones used in
our experiments were not in mint condition and some of the idiosyncrasies of individual microphones and speakers may have been the result of uneven
wear and tear on each device; we believe, however, that this is likely to occur in the real world as well.

\section{Conclusion}{\label{conclusion}}

In this paper we show that it is feasible to fingerprint smart devices through on-board acoustic components like microphones and speakers. As
microphones and speakers are one of the most standard components present in almost all smart devices available today, this creates a key privacy
concern for users. By the same token, efficient fingerprinting may also serve to enhance authentication. To demonstrate feasibility of this approach,
we collect fingerprints from five different brands of smartphones, as well as from 15 smartphones manufactured by the same vendor. Our studies show
that it is possible to successfully fingerprint smartphones through microphones and speakers, not only under controlled environments, but also in the
presence of ambient noise. We believe our findings are important steps towards understanding the full consequences of fingerprinting smart devices
through acoustic channels.
\section*{Acknowledgement}
We would like to thank Thomas S. Benjamin for his valuable input during the initial phase of the project. We would also like to thank the Computer Science department at UIUC and Google for providing us with the Motorola Droid phones.

{\footnotesize
\bibliographystyle{acm}
\bibliography{bibliograph}
}

\appendix
\section{Audio Features}{\label{appendix}}

\paragraphb{Root-Mean-Square (RMS) Energy:}
This feature computes the square root of the arithmetic mean of the squares of the original audio signal strength at various frequencies. In the case
of a
set of $N$ values $\{x_1,x_2,\dots,x_N\}$, the RMS value is given by the following formula:\nolinebreak
\begin{equation}
x_{\mathrm{rms}} = \sqrt{\frac{1}{n} \left( x_1^2 + x_2^2 +\cdots + x_N^2 \right)}
\end{equation}
The RMS value provides an approximation of the average audio signal strength.

\paragraphb{Zero Crossing Rate (ZCR):}
The zero-crossing rate is the rate at which the signal changes sign from positive to negative or back~\cite{chen1988signal}. This feature has been
used heavily in both speech recognition and music information retrieval, for example to classify percussive sounds~\cite{Gouyon00onthe}. ZCR for a
signal $s$ of length $T$ can be defined as:\nolinebreak
\begin{equation}
 ZCR = \frac{1}{T}\sum_{t=1}^{T}|s(t)-s(t-1)|
\end{equation}
where $s(t)=1$ if the signal has a positive amplitude at time $t$ and 0 otherwise. Zero-crossing rates provide a measure of the noisiness of the
signal.

\paragraphb{Low Energy Rate:}
The low energy rate computes the percentage of frames (typically 50ms chunks) with RMS power less than the average RMS power for the whole audio
signal. For instance, a musical excerpt with some very loud frames and a lots of silent frames would have a high low-energy rate.

\paragraphb{Spectral Centroid:}
The spectral centroid represents the ``center of mass'' of a spectral power distribution. It is calculated as the weighted mean of the frequencies
present in the signal, determined using a fourier transform, with their magnitudes as the weights:\nolinebreak
\begin{equation}
Centroid, \mu= \frac{\sum_{i=1}^{N} f_i \cdot m_i } {\sum_{i=1}^{N}m_i }\label{centroid}
\end{equation}
where $m_i$ represents the magnitude of bin number $i$, and $f_i$ represents the center frequency of that bin.

\paragraphb{Spectral Entropy:}
Spectral entropy captures the spikiness of a spectral distribution. As a result spectral entropy can be used to capture the formants or peaks in the
sound envelope~\cite{Misra2004}. To compute spectral entropy, a Digital Fourier Transform (DFT) of the signal is first carried out. Next, the
frequency
spectrum is converted into a probability mass function (PMF) by normalizing the spectrum using the following equation:\nolinebreak
\begin{equation}
w_i=\frac{m_i}{\sum_{i=1}^N m_i}\label{pmf}
\end{equation}
where $m_i$ represents the energy/magnitude of the $i$-th frequency component of the spectrum, $w=(w_1,w_2,\dots,w_N)$
is the PMF of the spectrum and N is the number of points in the spectrum. This PMF can then be used to compute the
spectral entropy using the following equation:\nolinebreak
\begin{equation}
H=\sum_{i=1}^N w_i \cdot log_{2}w_i
\end{equation}
The central idea of using entropy as a feature is to capture the peaks of the spectrum and their location.

\paragraphb{Spectral Irregularity:}
Spectral irregularity measures the degree of variation of the successive peaks of a spectrum. This feature provides the ability to capture the jitter
or noise in
spectra. Spectral irregularity is computed as the sum of the square of the difference in amplitude between adjoining spectral
peaks~\cite{jensen1999timbre} using the following equation:\nolinebreak
\begin{equation}
Irregularity = \frac{\sum_{i=1}^N (a_i - a_{i +1})^2 }{\sum_{i=1}^N a_i^2 }
\end{equation}
where the $(N+1)$-th peak is assumed to be zero. A change in irregularity changes the perceived timbre of a sound.

\paragraphb{Spectral Spread:}
Spectral spread defines the dispersion of the spectrum around its centroid, i.e., it measures the standard deviation of the spectral distribution. So
it can be computed as:\nolinebreak
\begin{equation}
Spread, \sigma = \sqrt{\sum_{i=1}^{N}\left[(f_i-\mu)^2\cdot w_i\right]}\label{spread}
\end{equation}
where $w_i$ represents the weight of the $i$-th frequency component obtained from equation (\ref{pmf}) and $\mu$
represents the centroid of the spectrum obtained from equation (\ref{centroid}).

\paragraphb{Spectral Skewness:}
Spectral skewness computes the coefficient of skewness of a spectrum. Skewness (third central moment) measures the symmetry of the distribution. A
distribution can be positively skewed in which case it has a long tail to the right while a negatively-skewed distribution has a longer tail to the
left. A symmetrical distribution has a skewness of zero. The coefficient of skewness is the ratio of the skewness to the standard deviation raised to
the third power.\nolinebreak
\begin{equation}
Skewness = \frac{\sum_{i=1}^{N}\left[(f_i-\mu)^3 \cdot w_i\right]}{\sigma^3}
\end{equation}

\paragraphb{Spectral Kurtosis:}
Spectral Kurtosis gives a measure of the flatness or spikiness of a distribution relative to a normal distribution. It
is computed from the fourth central moment using the following function:\nolinebreak
\begin{equation}
Kurtosis = \frac{\sum_{i=1}^{N}\left[(f_i-\mu)^4 \cdot w_i\right]}{\sigma^4}
\end{equation}
A kurtosis value of 3 means the distribution is similar to a normal distribution whereas values less than 3 refer to flatter distributions and values
greater than 3 refers to steeper distributions.

\paragraphb{Spectral  Rolloff:}
The spectral rolloff is defined as the frequency below which 85\% of the distribution magnitude is concentrated~\cite{Tzanetakis2002}
\begin{equation}
\argmin_{f_c\in\{1,\dots,N\}}\sum_{i=1}^{f_c}m_i \geq 0.85\cdot\sum_{i=1}^{N}m_i
\end{equation}
where $f_c$ is the rolloff frequency and $m_i$ is the magnitude of the $i$-th frequency component of the spectrum. The
rolloff is another measure of spectral shape that is correlated to the noise cutting frequency~\cite{Peeters2004}.

\paragraphb{Spectral Brightness:}
Spectral brightness calculates the amount of spectral energy corresponding to frequencies higher than a given cut-off threshold. This metric
correlates to the perceived timbre of a sound. Increase of higher frequency energy in the spectrum yields a sharper timbre, whereas a decrease yields
a softer timbre~\cite{Juslin64224}. Spectral brightness can be computed using the following equation:
\begin{equation}
Brightness_{f_c} = \sum_{i=f_c}^{N}m_i
\end{equation}
where $f_c$ is the cut-off frequency (set to 1500Hz) and $m_i$ is the magnitude of the $i$-th frequency component of the
spectrum.

\paragraphb{Spectral Flatness:}
Spectral flatness measures how energy is spread across the spectrum, giving a high value when energy is equally distributed and a low value when
energy is concentrated in a small number of narrow frequency bands. The spectral flatness is calculated by dividing the geometric mean of the power
spectrum by the arithmetic mean of the power spectrum~\cite{Johnston88}:\nolinebreak
\begin{equation}
Flatness=\frac {\left[\prod_{i=1}^N m_i\right]^{1/N}} {\frac{1}{N}\sum_{i=1}^N m_i}
\end{equation}
where $m_i$ represents the magnitude of bin number $i$. Spectral flatness provides a way to quantify the noise-like or tone-like nature of the
signal. One advantage of using spectral flatness is that it is not affected by the amplitude of the signal, meaning spectral flatness virtually
remains unchanged when the distance between the sound source and microphone fluctuates during recording.

\begin{figure}[!htb]
\centering
\includegraphics[width=0.7\columnwidth]{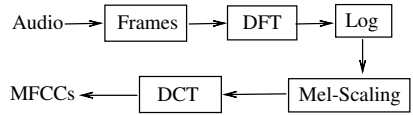}
\caption{Procedure for extracting MFCCs from audio signals} \label{mfcc}
\end{figure}

\paragraphb{Mel-Frequency Cepstrum Coefficients (MFCCs):}
MFCCs are short-term spectral features and are widely used in the area of audio and speech processing~\cite{Logan00melfrequency,Tzanetakis2002}.
Their success has been due to their capability of compactly representing spectrum amplitudes. Figure~\ref{mfcc} highlights the procedure for
extracting MFCCs from audio signals. The first step is to divide the signal into fixed size frames (typically 50ms chunks) by applying a windowing
function at fixed intervals. The next step is to take Discrete Fourier Transform (DFT) of each frame. After taking the log-amplitude of the magnitude
spectrum, the DFT bins are grouped and smoothed according to the perceptually motivated Mel-frequency scaling\footnote{Mel-scale approximates the
human auditory response more closely than the linearly-spaced frequency bands. \url{http://en.wikipedia.org/wiki/Mel_scale}}. Finally, in order to
decorrelate the resulting feature vectors a discrete cosine transform is performed. We use the first 13 coefficients for our experiments.

\paragraphb{Chromagram:}
A chromagram (also known as harmonic pitch class profile) is a 12-dimensional vector representation of an audio signal showing the distribution of
energy along the 12 distinct semitones or pitch classes. First a DFT of the audio signal is taken and then the spectral frequencies are mapped onto a
limited set of 12 chroma values in a many-to-one fashion~\cite{Fujishima99}. In general, chromagrams are robust to noise (e.g., ambient noise or
percussive sounds) and independent of timbre change.

\paragraphb{Tonal Centroid:}
Tonal centroid introduced by Harte et al.~\cite{Harte:2006} maps a chromagram onto a six-dimensional Hypertorus structure. The resulting
representation
wraps around the surface of a Hypertorus, and can be visualized as a set of three circles of harmonic pitch intervals: fifths, major thirds, and
minor thirds. Tonal centroids are efficient in detecting changes in harmonic contents.

\end{document}